\documentclass[iicol,pdflatex,sn-mathphys-num]{sn-jnl}

\usepackage{graphicx}%
\usepackage{multirow}%
\usepackage{amsmath,amssymb,amsfonts}%
\usepackage{amsthm}%
\usepackage{mathrsfs}%
\usepackage[title]{appendix}%
\usepackage{xcolor}%
\usepackage{textcomp}%
\usepackage{manyfoot}%
\usepackage{booktabs}%
\usepackage{algorithm}%
\usepackage{algorithmicx}%
\usepackage{algpseudocode}%
\usepackage{listings}%
\usepackage{caption}
\usepackage{subcaption}

\newcommand{\Rekk}{\ensuremath{Re_{kk}}}
\newcommand{\ws}{{\color{white}$\star\,$}}
\newcommand{\bs}{{$\star\,$}}
\newcommand{\wz}{{\color{white}0}}
\newcommand{\xo}{\ensuremath{\overline{x}_0}}
\newcommand{\xbar}{\ensuremath{\overline{x}}}

\begin{document}

\title[CHANGE Roughness-induced transition and turbulent wedge spreading]{Roughness-induced transition and turbulent wedge spreading}


\author*[1]{\fnm{Alexandre} \sur{Berger}}\email{aberger2@fsu.edu}

\author[2]{\fnm{Edward} \sur{White}}\email{ebw@tamu.edu}
\equalcont{These authors contributed equally to this work.}

\affil*[1]{\orgdiv{Department of Mechanical Engineering}, \orgname{Florida State University}, \orgaddress{\city{Tallahassee}, \postcode{32310}, \state{FL}, \country{USA}}}

\affil[2]{\orgdiv{Department of Aerospace Engineering}, \orgname{Texas A\&M University}, \orgaddress{\city{College Station}, \postcode{77834}, \state{TX}, \country{USA}}}

\abstract{Boundary layer transition triggered by a discrete roughness element generates a turbulent wedge that spreads laterally as it proceeds downstream. The historical literature reports the spreading half angle is approximately 6$^{\circ}$ in zero-pressure gradient flows regardless of Reynolds number and roughness shape. Recent simulations and experiments have sought to explain the lateral-spreading mechanism and have observed high- and low-speed streaks along the flanks of the wedge that appear central to the spreading process. To better elucidate the roles of Reynolds number and of streaks, a naphthalene flow-visualization survey and hotwire measurements are conducted over a wider range of Reynolds numbers and longer streamwise domain than previous experiments. The naphthalene results show that, while the mean spreading angle is consistent with the historical literature,there may be a weak dependency on $x$-based Reynolds number, which emerges as a result of the large sample size of the survey. The distance between the roughness element and the wedge origin exhibits a clear trend with the roughness-height-based Reynolds number. The hotwire measurements explain that this difference originates from whether breakdown occurs first in the central lobe or flanking streaks of the turbulent wedge. This observation highlights different transition dynamics at play within the supercritical regime. In agreement with past experiments, the hotwire measurements reveal that breakdown occurs in the wall normal shear layer above low-speed streaks. Due to the elongated streamwise extent of this experiment, secondary streak dynamics are also uncovered.  A high-speed streak is produced directly downstream of the initiating low-speed streak. Subsequently, a new low-speed streak is observed outboard of the previous high-speed streak. This self-sustaining process is the driving mechanism of turbulent wedge spreading.}




\maketitle

\section{Introduction}\label{sec1}

When boundary-layer transition is triggered by an isolated roughness element, a turbulent wedge is formed. The vertex of the wedge is located downstream of the initiating roughness and the edges of the wedge spread laterally at an essentially constant angle. While turbulent wedges have been studied for decades, the mechanisms through which they spread are not yet fully understood. Turbulent wedges were first studied by \citet{Gregory-ARC-56} and \citet{Schubauer-NACA-55} who observed that wedges are composed of a $6.4^{\circ}$ half angle turbulent core, surrounded by a $10.6^{\circ}$ half angle intermittent region. These spreading angles are remarkably constant throughout the literature and are only slightly sensitive to pressure gradient \citep{zhong}.

A wedge is initiated by a roughness element or similar feature that reorients incoming spanwise vorticity in the streamwise direction. This results in a pair of counter-rotating streamwise vortices that extend downstream of the roughness. (See \citet{Baker-JFM-79} and \citet{Rizzetta-AIAAJ-07} for experimental and numerical results, respectively.) These vortices redistribute streamwise momentum across the boundary layer and are responsible for the emergence of high- and low-speed streaks. \citet{Ergin-AIAAJ-06} observed that turbulence results from a Kelvin--Helmholtz-like instability of the shear layers located above the low-speed streaks.  This topology has been shown to be robust to changes in roughness geometry by \citet{Ye-IJHFF-16} using tomographic Particle-Image Velocimetry (PIV). Although some near-field details may be affected by roughness height and shape, the process ultimately leads to a turbulent wedge that undergoes lateral spreading.

To understand why turbulent wedges spread at the angle at which they do, \citet{GadElHak-JFM-81} used dye visualization in a water channel. That work revealed that lateral spreading due to turbulent mixing is much slower than what is observed for the overall lateral wedge spreading. This suggests that, in addition to turbulent mixing, there is a destabilization mechanism acting on the surrounding laminar flow which is responsible for the lateral growth of the wedge. \citet{Chu-AIAA-12} and \citet{Goldstein-FTC-17} used Direct Numerical Simulation (DNS) and confirmed that spreading is due to destabilization at the edge of the wedge rather than turbulent mixing from the core region. To establish this, they artificially damped turbulence in the core of the wedge and observed essentially no effect on lateral spreading. 

With or without core turbulence damped, the DNS shows a succession of stationary low-speed streaks along the edges of the wedge. \citet{Chu-AIAA-12} hypothesized these stationary streaks are pinned in place by the stationary streamwise vortices that originate from the roughness element and, subsequently, ``the spanwise growth of the wedge is the result of the formation of a spanwise succession of streaks, where the formation of each, in turn, is the result of an instability introduced by the vorticity field preceding it upstream.'' The succession of low-speed streaks has previously been observed in an experiment by \citet{Watmuff-AFMC-04} who generated a steady low-speed streak and excited it with high-frequency oscillations.  The streak broke down to turbulence and, where it did, a pair of laterally displaced low-speed streaks were generated. The initial low-speed streak plus the displaced secondary pair were associated with laterally displaced high-speed streaks farther downstream.  Based on this, Watmuff suggested that lateral spreading ``is the result of the formation of a spanwise succession of streaks, where the formation of each in turn is the result of an instability introduced by the streak immediately preceding it upstream.''

Inspired by the \citet{Chu-AIAA-12} DNS, \citet{Kuester-ExpFluids-16} undertook an experiment matched to the DNS conditions in an effort to observe that streaks are associated with lateral wedge spreading. Hotwire measurements paired with naphthalene shear stress visualization showed a pair of low-speed streaks generated by a discrete roughness element. These destabilized, became turbulent, and produced a new pair of low-speed streaks laterally displaced outboard. Somewhat downstream of the low-speed streaks, a pair of high-speed streaks appears further out from the symmetry line.

A complementary experimental view is provided by a tomographic PIV study by \citet{Ye-IJHFF-16}. Using measurements with $Re_{x,k} = 207\times 10^{3}$ and $Re_{kk} = 1170$, \citeauthor{Ye-IJHFF-16} observed small variations in wedge spreading angle, breakdown location, and streamwise streak spacing depending on roughness shape. $Re_{x,k}=U_{\infty}\,x_k/\nu$ is the Reynolds number based on the streamwise location of the roughness, $x_k$ \footnote{The $Re_{x,k}$ value is quoted for \citeauthor{Ye-IJHFF-16} is adjusted for the virtual leading edge using the $Re_{\theta}$ value at $x_k$ quoted by \citet{Ye-PhD-17}.}. $\Rekk=U(k)\,k/\nu$ is the Reynolds number based on roughness height, $k$, in which $U(k)$ is the velocity of the undisturbed laminar boundary layer at height $k$.

In \citeauthor{Ye-IJHFF-16}'s experiment, two pairs of counter-rotating streamwise vortices develop just behind the roughness and a low-speed region along the symmetry line is unstable to strong Kelvin--Helmholtz-like vortices. Outboard of the central features, trailing legs of a horseshoe vortex are associated with stationary streamwise vorticity that generate low-speed flow on their upwash sides similar to the low-speed streak observation by \citet{Kuester-ExpFluids-16}. Further downstream, high-speed streaks emerge further outboard where tertiary vortices produce downwash. Additional sets of streamwise vortices with alternating senses of rotation can also be observed. These measurements clearly show an alternating pattern of vortices lead to a repeating pattern of low-speed and high-speed streaks, exactly consistent with DNS results.

Hotwire measurements by \citet{Kuester-ExpFluids-16} and \citet{Berger-TSFP-17} give the frequency spectrum of velocity fluctuations at various locations in the flow. These confirm the findings by \citet{Ergin-AIAAJ-06} that roughness-induced transition occurs due to a high-frequency instability above and wrapping around the sides of low-speed streaks. These fluctuations are clearly associated with the unsteady vortices observed by \citet{Ye-IJHFF-16} and are similar to those above the low-speed streaks in various DNS studies. To establish that turbulent wedge spreading results from an instability process, \citet{Berger-TSFP-17} performed a linear-stability analysis of a two-dimensional slice of a low-speed streak as a basic state and found the most-unstable frequencies in good agreement with the measured fluctuation spectrum.

The long-standing observations of constant wedge spreading angles suggests a universal behavior. However, few experiments provide details about the mechanisms that drive lateral spreading. Previous experiments using roughness-initiated turbulent wedges have been performed at fixed and relatively low $Re_{x,k}$ and over relatively short streamwise distances. This work addresses the need for data on wedge-spreading mechanisms and universality over multiple wedge realizations, an extended downstream domain, and across a wider range of Reynolds numbers than have been explored previously. It additionally investigates the emergence of high- and low-speed streaks along the sides of multiple realizations of roughness-induced turbulent wedges. Naphthalene shear stress visualizations are made for 17 separate turbulent wedges initiated at $Re_{x,k}$ values from $231\times 10^{3}$ to $1.505\times 10^6$, which span from the Reynolds numbers tested by \citet{Ye-IJHFF-16}, $Re_{x,k}=207\times 10^{3}$, to almost three times higher than the values tested by \citet{Kuester-ExpFluids-16}, $Re_{x,k}=541\times 10^{3}$. The measurement domain extends at least $78\,\delta^*_k$ and up to $265\,\delta^*_k$. The work by \citeauthor{Ye-IJHFF-16} and \citeauthor{Kuester-ExpFluids-16} extended only from $116\,\delta^*_k$ and $132\,\delta^*_k$ from the roughness element. Extensive hotwire measurements of mean and fluctuating streamwise velocity are made at each of the $Re_{kk}$ values. Of particular interest is the growth of fluctuations expected above the low-speed streak.

\section{Experimental Conditions}\label{sec2}

The experiments are performed using a flat plate model in the Klebanoff--Saric Wind Tunnel (KSWT) at Texas A\&M University. The KSWT is a low-speed wind tunnel designed to produce minimal freestream turbulence and sound. The test section is 1.4 m $\times$ 1.4 m and is 5 m long. The freestream turbulence intensity is $Tu = 0.02\%$ (DC to 10~kHz) and acoustic fluctuations are less than 85~dB above 30~Hz \cite{Hunt-AIAA-10,Kuester-AIAA-12}.

The flat plate spans the height of the test section and is 4.4~m long including a 343-mm-long elliptic leading edge. Zero pressure gradient is confirmed using boundary-layer profile measurements at each operating unit Reynolds number. Measured shape factors $H=2.59\pm0.04$ indicate the pressure gradient is suitably close to zero for stability experiments \citep{Saric-SHEFM-07}. The same measurements provide the virtual leading edge, $x_{\mathrm{vle}}$, of the Blasius boundary layer such that $\delta^* = 1.72 [(x-x_{\mathrm{vle}})/Re']^{1/2}$ where $x$ is the actual distance from the leading edge and $Re'=U_{\infty}/\nu$ is the unit Reynolds number.

The experiments study turbulent wedges generated by two cylindrical roughness elements of the same design used by \citet{Kuester-ExpFluids-16} with $k/d=1.05$. The larger roughness is $k=2.94$ mm and $d=2.79$; the smaller is $k=1.96$ mm and $d=1.86$. The roughness elements both feature a 0.2~mm $\times$ 0.2~mm square notch on the $+z$ side to intentionally break symmetry for matched DNS studies \cite[e.g.,][]{Chu-AIAA-12}.

To provide a range of  $Re_{x,k}$ values at the fixed $Re_{kk}$ values, the roughness elements are located at various streamwise positions while the wind tunnel is operated at different unit Reynolds numbers. Streamwise locations are identified for the two roughness elements that provide the three $Re_{kk}$ values: 600, 750, and 979. Table~\ref{condition-table} lists the 17~separate combinations of roughness position, size, Reynolds number, and $\delta^*_k$ values included in the naphthalene visualization study. Also listed are the lengths, $L$, of the useful measurement range downstream of the element. The streamwise positions of the roughnesses are relative to the physical leading edge of the plate. The virtual leading edge lengths are accounted for in the Reynolds numbers and displacement thicknesses.

\begin{table*}
    \caption{Summary of naphthalene study test conditions. Conditions for
    hotwire measurements are marked by stars.}
    \begin{tabular}{*{11}{c}c} \hline\hline

$Re_{kk}$ & $Re_{x,k}$ & $Re'$ [$\mathrm{m}^{-1}$] & $x_k$ [mm] &
$\delta^*_k$ [mm] & $k$ [mm] & $L$ [mm] & $k/\delta^*_k$ & $L/\delta^*_k$ \\[2mm] \hline

\ws600 & $\wz\,265\!\times\!10^3$ & $508\!\times\!10^3$ & \wz\,718 & 1.74 & 1.96 & 376 & 1.12 & 215 \\ 
\ws600 & $\wz\,380\!\times\!10^3$ & $550\!\times\!10^3$ & \wz\,878 & 1.93 & 1.96 & 512 & 1.02 & 265 \\
\ws600 & $\wz\,554\!\times\!10^3$ & $600\!\times\!10^3$ & 1\,102   & 2.14 & 1.96 & 425 & 0.92 & 199 \\
\bs600 & $\wz\,778\!\times\!10^3$ & $650\!\times\!10^3$ & 1\,366   & 2.33 & 1.96 & 281 & 0.84 & 120 \\
\ws600 & $1\,059\!\times\!10^3$   & $700\!\times\!10^3$ & 1\,676   & 2.53 & 1.96 & 553 & 0.78 & 219 \\
\ws600 & $1\,407\!\times\!10^3$   & $750\!\times\!10^3$ & 2\,033   & 2.72 & 1.96 & 347 & 0.72 & 128 \\[2mm]

\ws750 & $\wz\,231\!\times\!10^3$ & $550\!\times\!10^3$ & \wz\,607 & 1.50 & 1.96 & 326 & 1.30 & 217 \\
\ws750 & $\wz\,343\!\times\!10^3$ & $600\!\times\!10^3$ & \wz\,750 & 1.68 & 1.96 & 345 & 1.17 & 205 \\
\ws750 & $\wz\,488\!\times\!10^3$ & $650\!\times\!10^3$ & \wz\,920 & 1.85 & 1.96 & 307 & 1.06 & 166 \\
\ws750 & $\wz\,712\!\times\!10^3$ & $700\!\times\!10^3$ &   1\,180 & 2.07 & 1.96 & 261 & 0.95 & 126 \\
\bs750 & $\wz\,892\!\times\!10^3$ & $750\!\times\!10^3$ &   1\,346 & 2.17 & 1.96 & 265 & 0.90 & 122 \\
\ws750 &   $1\,165\!\times\!10^3$ & $800\!\times\!10^3$ &   1\,607 & 2.32 & 1.96 & 265 & 0.84 & 114 \\
\ws750 &   $1\,493\!\times\!10^3$ & $850\!\times\!10^3$ &   1\,902 & 2.47 & 1.96 & 272 & 0.79 & 110 \\[2mm] 

\bs979 & $\wz\,523\!\times\!10^3$ & $508\!\times\!10^3$ & 1\,255 & 2.45 & 2.94 & 500 & 1.20 & 204 \\ 
\ws979 & $\wz\,752\!\times\!10^3$ & $550\!\times\!10^3$ & 1\,555 & 2.71 & 2.94 & 483 & 1.08 & 178 \\
\ws979 &   $1\,082\!\times\!10^3$ & $600\!\times\!10^3$ & 1\,981 & 2.98 & 2.94 & 294 & 0.99 & \wz\,99 \\
\ws979 &   $1\,505\!\times\!10^3$ & $650\!\times\!10^3$ & 2\,485 & 3.25 & 2.94 & 253 & 0.91 & \wz\,78 \\ \hline\hline

\end{tabular}
    \label{condition-table}
\end{table*}

\section{Naphthalene Wedge Geometry Survey}

In order to efficiently evaluate wedge spreading over 17~distinct wedges, naphthalene shear-stress visualization is used to rapidly detect the laminar/turbulent interface. The method consists of dissolving naphthalene in acetone and spraying the mixture onto the test surface. The acetone evaporates quickly and leaves behind a thin layer of milky white naphthalene crystals. When the tunnel is run, the naphthalene sublimates revealing the aluminum flat plate. Sublimation occurs more quickly in regions of higher shear stress so turbulent regions and high-speed laminar streaks are revealed as dark regions as opposed to the light-gray low-stress regions \cite{Dagenhart-AIAA-89,Kuester-ExpFluids-16,Berger-TSFP-17}. 

\begin{figure*}
    \centering
	\includegraphics[width=0.95\textwidth]{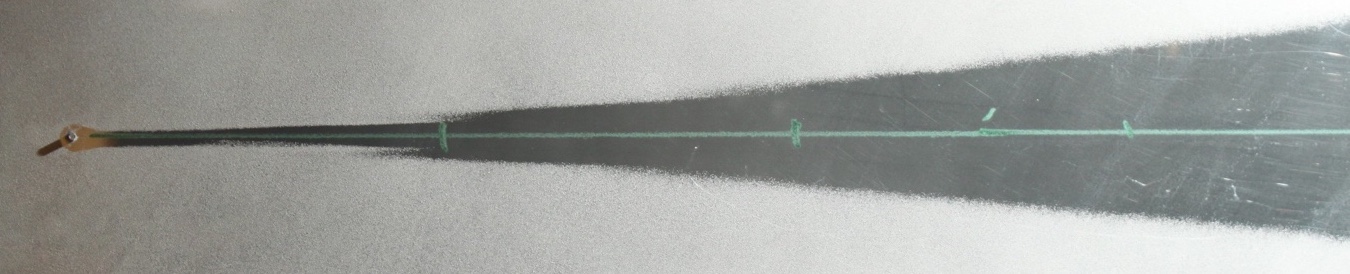}
	\caption{Example naphthalene image for $Re_{x,k}=343\times10^3$ and $Re_{kk} = 750$.} \label{fig:Rek750Wcr}
    \vspace{3mm}
	\includegraphics[width=0.95\textwidth]{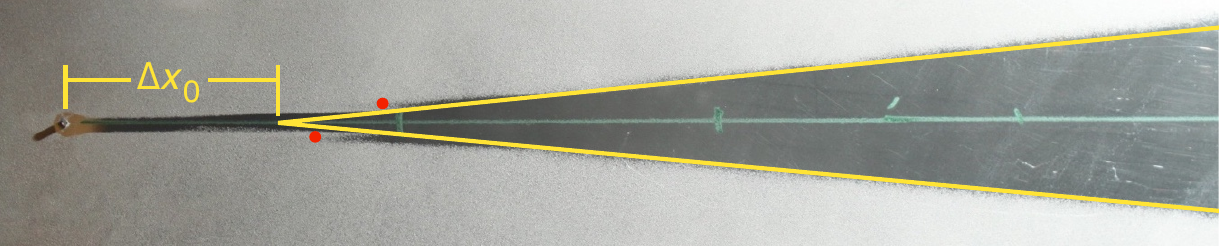}
	\caption{Wedge geometry analysis corresponding to
    figure~\ref{fig:Rek750Wcr}.}
    \label{fig:N600750}
\end{figure*}

As the experiment proceeds, multiple photographs, an example of which is provided in figure~\ref{fig:Rek750Wcr} reveal wedge boundaries as the naphthalene sublimates.  After image distortion is removed using a calibration image, geometry features are manually extracted. Wedge boundaries corresponding to figure~\ref{fig:Rek750Wcr} are shown in figure~\ref{fig:N600750}.  Their mutual angle is used to compute the wedge-spreading half angle. Their intersection is the effective wedge origin located $\Delta x_0$ downstream of the roughness element center. High-shear streaks marked as red dots in figure~\ref{fig:N600750} are often seen to originate outboard of the main wedge edge. These are not included in the spreading angle.

\begin{figure*}
    \centering
	\includegraphics[width=0.75\textwidth]{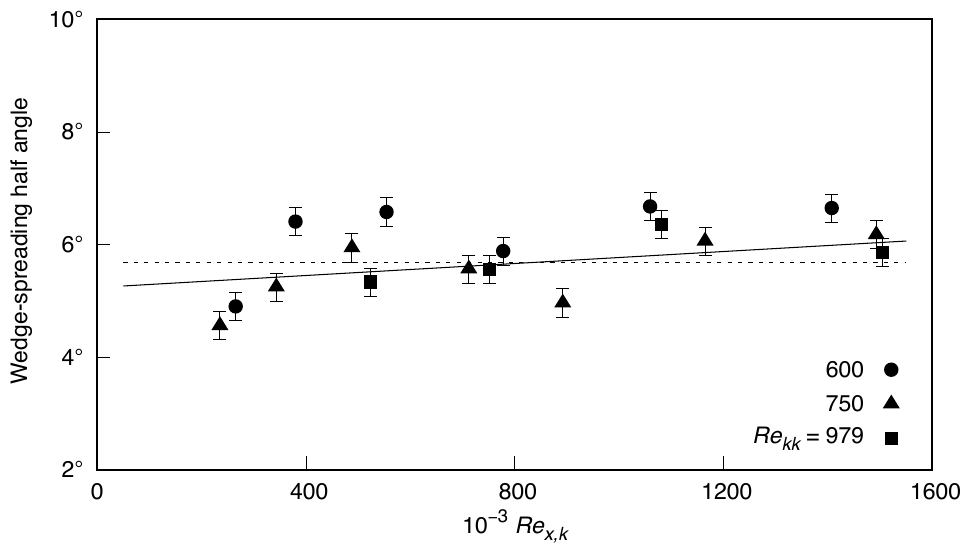}
	\caption{Turbulent wedge spreading half-angle. The dashed line is the mean, $5.8^{\circ}$. The solid line is the linear trend with $Re_{x,k}$.}
    \label{fig:spa}
\end{figure*}

Wedge-spreading half angles are plotted in figure~\ref{fig:spa}. The mean spreading half angle is $5.8^{\circ}$ with a standard deviation of $\pm 0.8^{\circ}$ and a standard error of the mean of $\pm 0.2^{\circ}$. The result $5.8^{\circ} \pm 0.2^{\circ}$ is representative of the turbulent core of the wedge and is in good agreement with earlier results reported by \citet{Schubauer-NACA-55}, \citet{zhong}, and \citet{Ye-IJHFF-16}.

Error bars on each point in figure~\ref{fig:spa} are $\pm0.25^{\circ}$ and represent the variability of manual image analysis across multiple naphthalene images. Spatial variability of the naphthalene coating also contributes to the uncertainty but is not possible to quantify. Including the measurement uncertainty in a linear model of spreading half angle versus $Re_{x,k}$ yields a weak dependence on Reynolds number, $0.81^{\circ}\pm0.15^{\circ}$ per million $Re_{x,k}$ with a mean spreading half angle of $5.8^{\circ} \pm 0.13^{\circ}$ at zero Reynolds number. This model matches the $5.7^{\circ}$ result near $Re_{x,k}=800\times10^3$. 

These results suggest that spreading half angle may be a weak function of Reynolds number but one that is difficult to confirm with certainty because of the data scatter. The present results provide the largest statistical sample of such results and over the largest range of $Re_{x,k}$ and are the first that are thought to suggest a potential variation in spreading angle with $Re_{x,k}$.

\begin{figure*}
    \centering
	\includegraphics[width=0.72\textwidth]{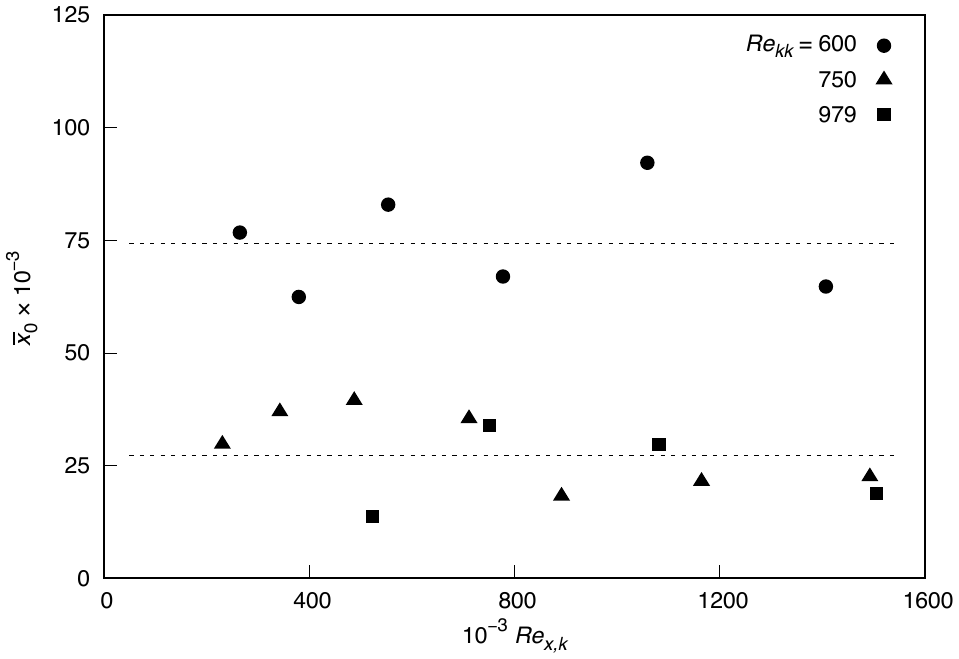}
	\caption{Nondimensional location of the turbulent wedge virtual origin, \xo.}
    \label{fig:dx0}
\end{figure*}

The distance from the roughness element to the virtual wedge origin, $\Delta x_0$, can be made nondimensional by $Re'$. To do so, a nondimensional streamwise distance is defined: $\overline{x}=Re'(x-x_k)$ with $\xo=Re'\Delta x_0$ denoting the nondimensional virtual wedge origin. Results in figure~\ref{fig:dx0} show no significant trend of across the range of $Re_{x,k}$ values tested. However, \xo\ does vary with $Re_{kk}$. The dashed lines in figure~\ref{fig:dx0} are mean values for $Re_{kk}=600$ and the combined $Re_{kk}=750$ and $979$ data. The two larger $Re_{kk}$ configurations have a mean of $\xo = 27\times10^3$ while the $Re_{kk}=600$ mean is about three times larger, $\xo = 74 \times10^3$. That is, the distance between the roughness element and the turbulent wedge origin is a function of $Re_{kk}$ but not $Re_{x,k}$ when properly nondimensionalized. This finding echoes the results by \citet{Klebanoff-JFM-92} that the point of breakdown to turbulence moves upstream toward the roughness element as $Re_{kk}$ is increased and that the distance to transition approaches an asymptotic minimum as $Re_{kk}$ increases.

\section{Hotwire Measurements and Results}

Hotwire measurements are made for three turbulent wedges initiated by $\Rekk=600$, 750, and~979 roughness elements as indicated in table~\ref{condition-table}. For each wedge, measurements are made in wall-normal/lateral planes at multiple $x$ locations each of which include between 21 and 81 boundary-layer profiles. To begin a boundary-layer profile measurement, a hotwire is positioned outside the boundary layer and moved toward the flat plate in progressively smaller steps, approximately 70 for laminar regions and 120 for turbulent regions. Individual points are sampled for 2~seconds at 10~kHz. Streamwise velocity data are separated into mean and fluctuating parts, $U(y)$ and $u'(y,t)$.

\subsection{Breakdown to turbulence}

\begin{figure*}
    \centering
    \includegraphics[width=0.98\textwidth]{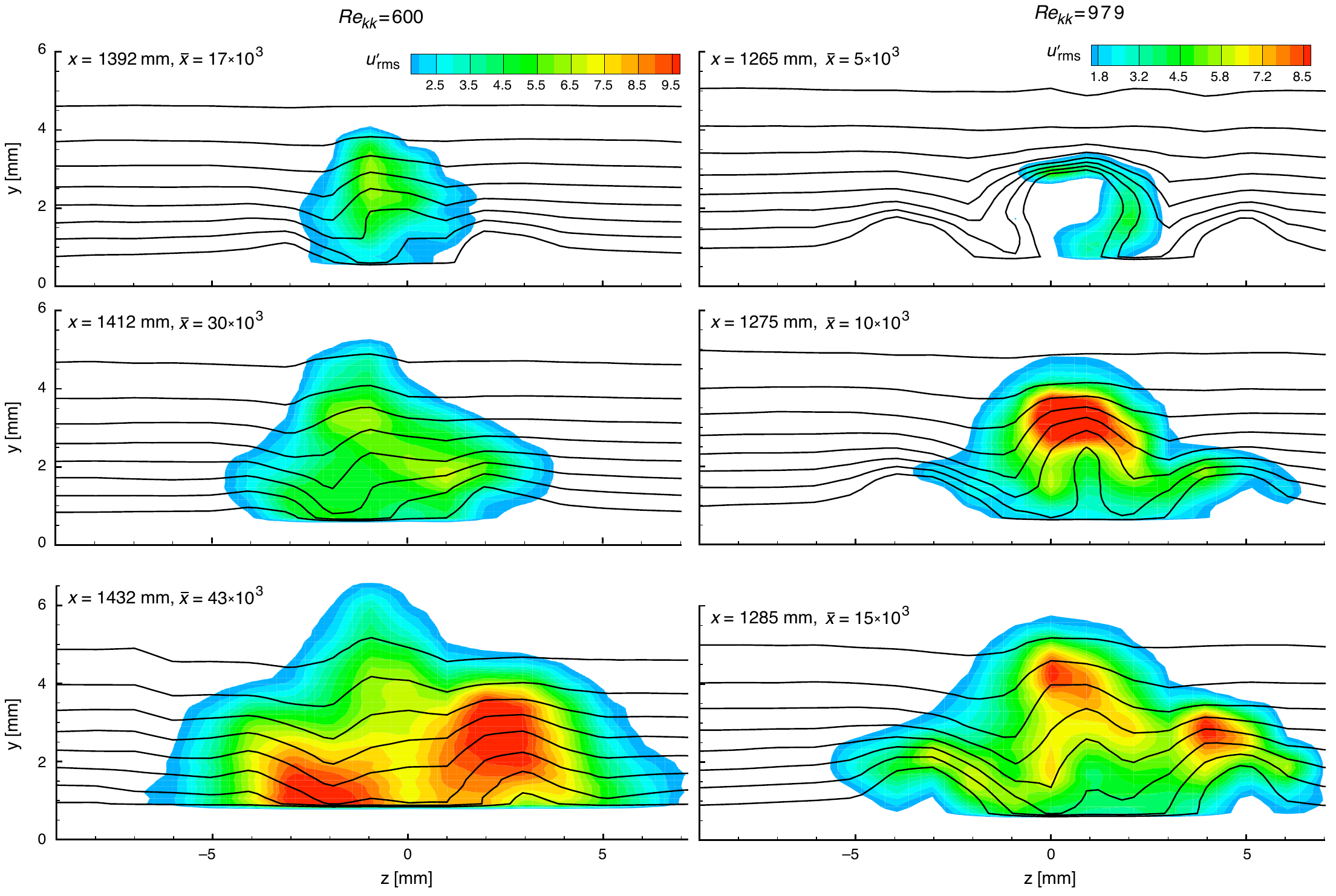}
    \caption{Measurements of the initial transition to turbulence behind
    $\Rekk=600$ (left) and~979 (right) roughness elements. Color
    contours indicate $u'_{\mathrm{rms}}/U_{\infty}$ levels. Contour lines
    are 10\% isolines of streamwise velocity between 20\% and 90\% of
    $U_{\infty}$. Contour plots in subsequent figures are plotted similarly.}
    \label{fig:Nearfield_Composite}
\end{figure*}

Naphthalene images show that the distance from the roughness to the virtual wedge origin is less for $\Rekk=979$ and~750 than for $\Rekk=600$ (figure~\ref{fig:dx0}).  Hotwire measurements in the near wakes of the roughness element reveal why. For the larger $\Rekk=979$ element, figure~\ref{fig:Nearfield_Composite} shows a lobe of low-velocity fluid high in the boundary layer on the $z=0$ centerline directly aft of the roughness element.  At the top of this feature, strong shear supports high-intensity $u'$ fluctuations that rapidly break down to turbulence.  $\Rekk = 600$ results in figure~\ref{fig:Nearfield_Composite} do not show this behavior. Instead, breakdown occurs first just above flanking low-speed streaks somewhat farther downstream. Equivalent plots for $\Rekk=750$ would resemble the $\Rekk=979$ results. The difference between these patterns appears to be responsible for the different \xo\ values described in figure~\ref{fig:dx0} and likely explains the results by \citet{Klebanoff-JFM-92}.

\subsection{Wedge spreading}

\begin{figure*}
    \centering
    \includegraphics[width=0.75\textwidth]{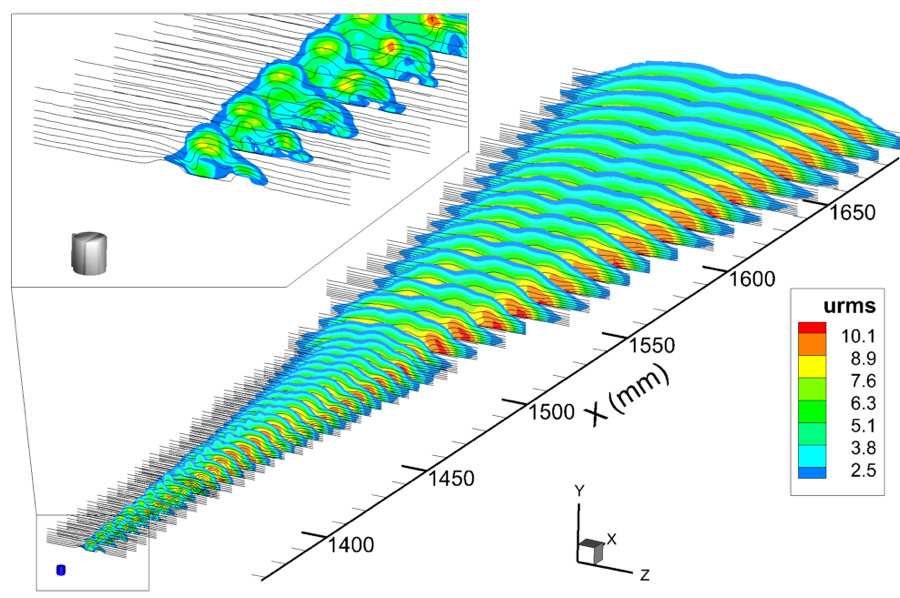}
    \caption{Measurements of the $\Rekk = 750$ turbulent wedge.}
    \label{fig:O750}
\end{figure*}

Composite results for the most-extensive hotwire data set, $\Rekk = 750$, are shown in figure~\ref{fig:O750}. For this configuration, the flow first breaks down to turbulence on the $z=0$ centerline at $x=1370$~mm, 24~mm behind the roughness, or $\xbar=18\times10^3$. Initial centerline breakdown is followed by breakdown just above the low-speed flanking streaks.

\begin{figure*}
    \centering \includegraphics[width=0.80\textwidth]{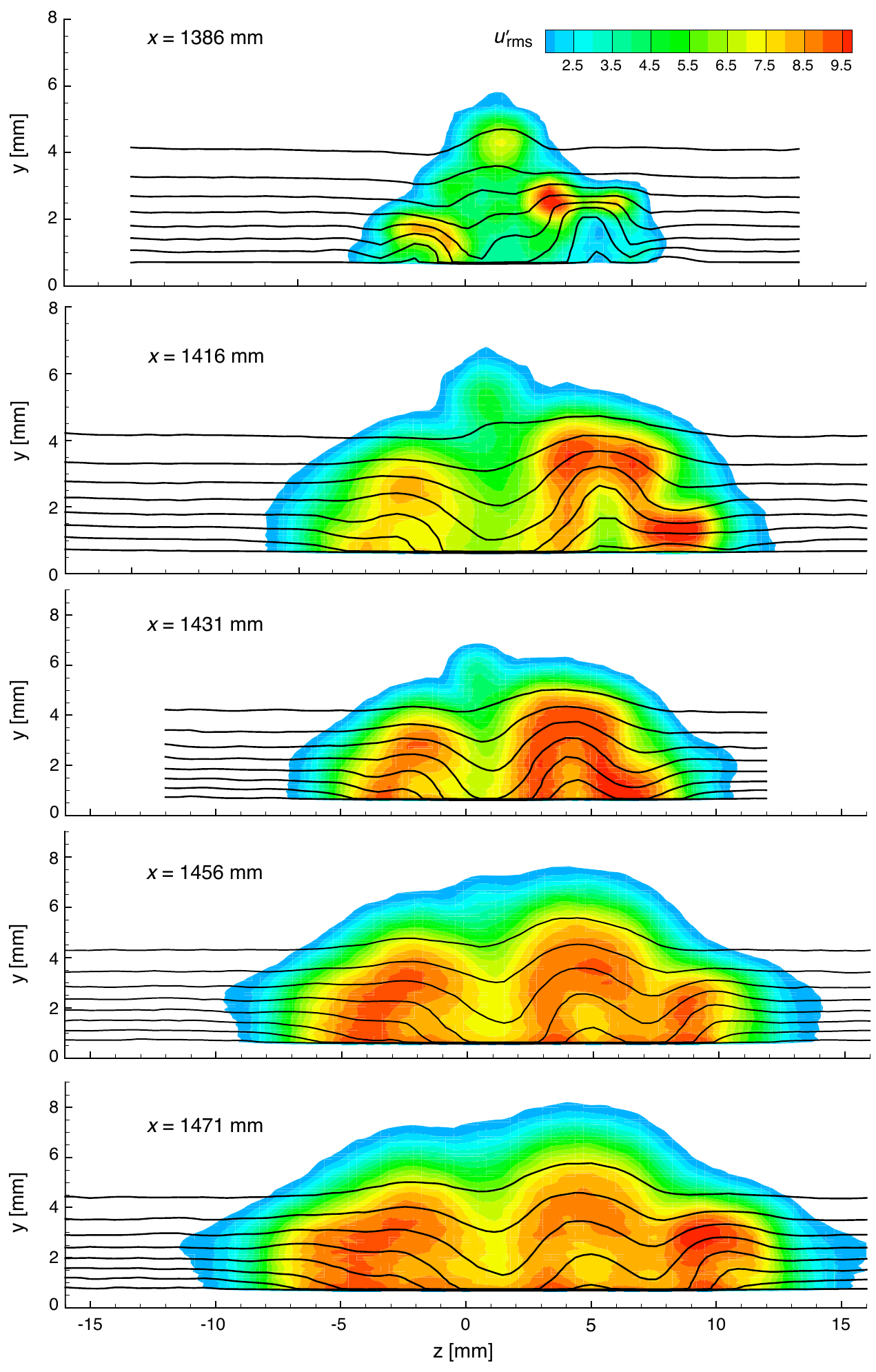}
    \caption{Breakdown of initial flanking streaks in the $\Rekk=750$ roughness wake.}
    \label{fig:Evolution2Rek750}
\end{figure*}

The detailed process by which flanking streaks become turbulent is presented in figure~\ref{fig:Evolution2Rek750}.  The top two plots show the initial low-speed streaks that result directly from the $\Rekk 750$ roughness element. The bottom three show the emergence and breakdown of a second set of flanking streaks that arise via the lateral wedge-spreading process. Earlier hotwire measurements by \citet{Kuester-ExpFluids-16} and \citet{Berger-TSFP-17} did not observe the emergence and breakdown of a second set of low-speed streaks. Thus, this is thought to be the first observation of flanking streak breakdown using hotwire anemometry that provides well-resolved fluctuation data.

Break down to turbulence on the centerline has already occurred upstream of the plots in figure~\ref{fig:Evolution2Rek750} and there is high wall shear at $z=0$. The wedge is not perfectly symmetric and the description here focuses on the $+z$ side where the behaviors are more clear. Data at $x=1386$~mm, $\xbar=30\times10^3$, shows that the low-speed streak near $z=+4$~mm is very pronounced with strong shear, especially $\partial U/\partial y$ above the streak. Moving downstream, the shear relaxes but the upper region continues to support growing $u'$ fluctuations that eventually wrap around the entire low-speed streak. At the same position, a new high-speed streak is forming outboard, around $z=+7$ or +8~mm.  Meanwhile, inboard of the low-speed streak, velocity contours are plunging toward the wall indicating high wall shear and emergence of turbulence at that $z$ location.

Further downstream at $\xbar=64\times10^3$, there is a more-developed high-speed streak with high-intensity $u'$ fluctuations at $z=+7$~mm. At $\xbar=83\times10^3$, $U$~iso-lines begin to lift up around $z=+10$~mm and form a new low-speed streak outboard of the high-speed region.  At this outboard location, the iso-lines indicate the presence of a streamwise vortex with counterclockwise rotation that corresponds exactly to the location and rotation of the ``fourth pair'' streamwise vortex observed by \citet{Ye-IJHFF-16} (figure~5a) using tomographic PIV. Then, at $\xbar=93\times10^3$, high intensity $u'$ fluctuations appear above the flanking low-speed streak in the region of elevated $\partial U/\partial y$ shear.  This suggests a continued pattern of high- and low-speed streaks that emerge and break down to cause lateral wedge spreading.  The same process of emergence and breakdown flanking streaks is seen in the $\Rekk=600$ and~$979$ wedge data. 

\begin{figure*}
    \centering \includegraphics[width=0.80\textwidth]{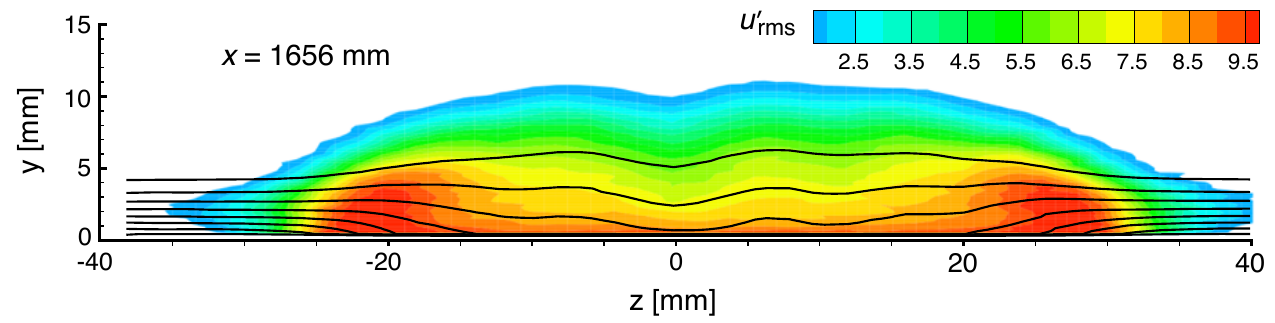}
    \caption{Measurements of the $\Rekk = 750$ wedge far downstream of the roughness element.}
    \label{fig:Evolution3Rek750}
\end{figure*}

Much farther downstream, figure~\ref{fig:Evolution3Rek750} shows that the strongest $u'$ levels persist near the wall at the edges of the turbulent core. A distinct overhang region with lower-intensity fluctuations extends beyond the turbulent core of the wedge. The overhang is likely to be what was reported by \citet{Schubauer-NACA-55} as the $10.6^{\circ}$ intermittent region. The present hotwire measurements show a $6.6^{\circ}$ spreading half angle based on locations where the mean velocity is elevated close to the wall and rms velocity fluctuations are largest.  The corresponding naphthalene image yields a $6.8^{\circ}$ half angle which further emphasizes that naphthalene images reveal the turbulent core.

In the far downstream region, it is difficult to distinguish clear streak structures along the flanks of the wedge. It is possible the high- and low-speed streak cycle persists in the downstream region but simply cannot be detected in time-averaged data. Data analyzed by \citet{Berger-PhD-20} suggest the streak meanders in the lateral direction and, by doing so, obscures the distinct streak structures in the time-averaged hotwire signals. A carefully designed conditional sampling approach that builds on Berger's meandering concept would be required to definitively confirm this hypothesis. 

\subsection{Velocity fluctuation power spectra}

\begin{figure*}
    \centering \includegraphics[width=0.85\textwidth]{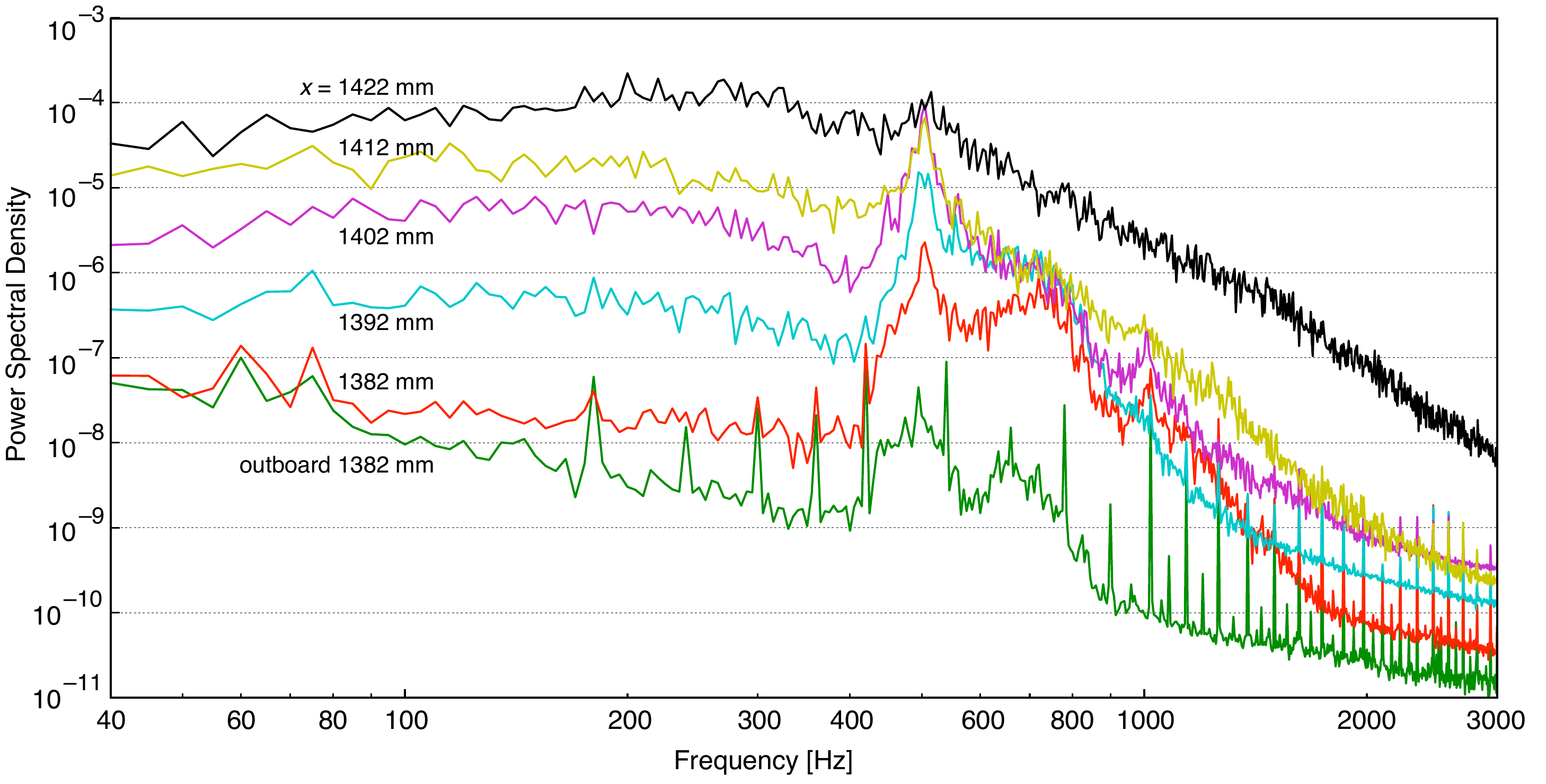}
    \caption{Velocity fluctuation power spectra in the initial low-speed streak for $\Rekk = 600$. Spectra are color-coded corresponding to the locations indicated in figure~\ref{fig:locPS6e}.}
    \label{fig:Rek600PSDPPS} 
\end{figure*}

\begin{figure}
    \centering \includegraphics[width=0.45\textwidth]{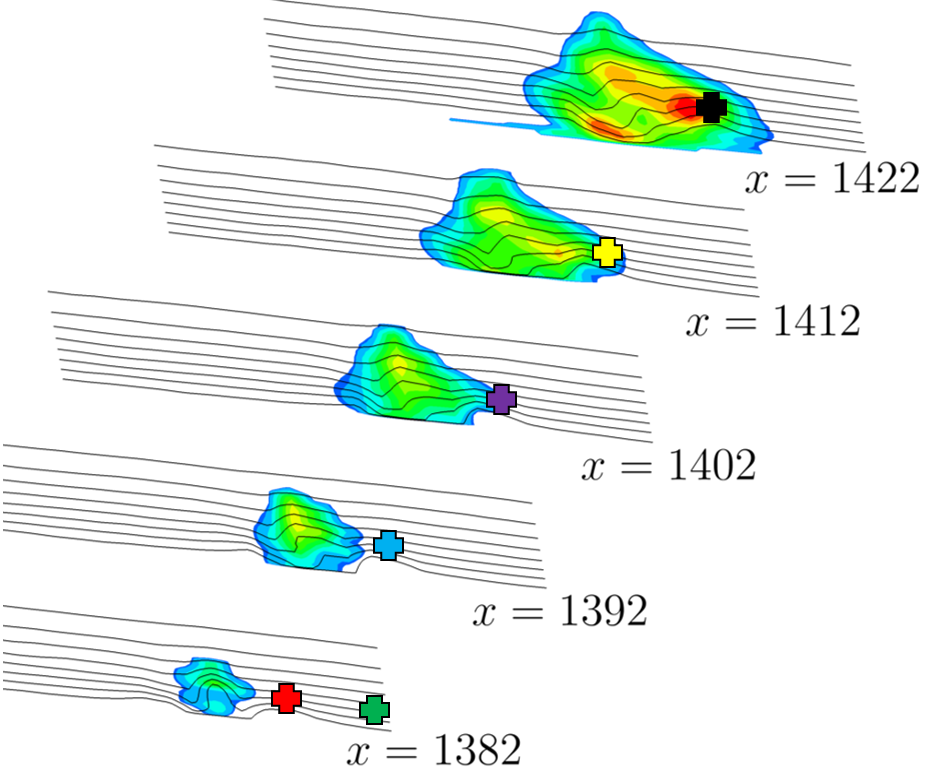}
    \caption{Locations of power spectra shown in figure~\ref{fig:Rek600PSDPPS}.} \label{fig:locPS6e}
\end{figure}

An advantage of hotwire anemometry is the ability to measure well-resolved $u'$ spectra that yield unstable frequencies. Observations here focus on the region directly above the flanking low-speed streaks where fluctuations are associated with the wedge-spreading streak cycle. Fluctuation spectra for $\Rekk = 600$ are given in figure~\ref{fig:Rek600PSDPPS} corresponding to measurement locations shown in figure~\ref{fig:locPS6e}. At the most-upstream portion of the streak at $x=1382$~mm, $\xbar=10\times10^3$, there is a band of unstable frequencies from 400~to 800~Hz. Corresponding dimensionless frequencies are $F = 2\pi f\nu/U_{\infty}^2 = 375$ to $750\times10^{-6}$, much higher than unstable Tollmien--Schlichting frequencies. Moving downstream, the streak becomes more pronounced and there is rapid growth of the unstable band of frequencies, especially near 500~Hz. By $x=1422$~mm, $\xbar=36\times10^{3}$, the flow has become locally turbulent with a broad $u'$ spectrum. Similar to the observations by \citet{Ergin-AIAAJ-06}, \citet{Kuester-ExpFluids-16}, and \citet{Berger-TSFP-17}, these fluctuations appear to be due to a Kelvin--Helmholtz-like shear-layer instability where gradients of streamwise velocity are large.

\begin{figure*}
    \centering \includegraphics[width=0.85\textwidth]{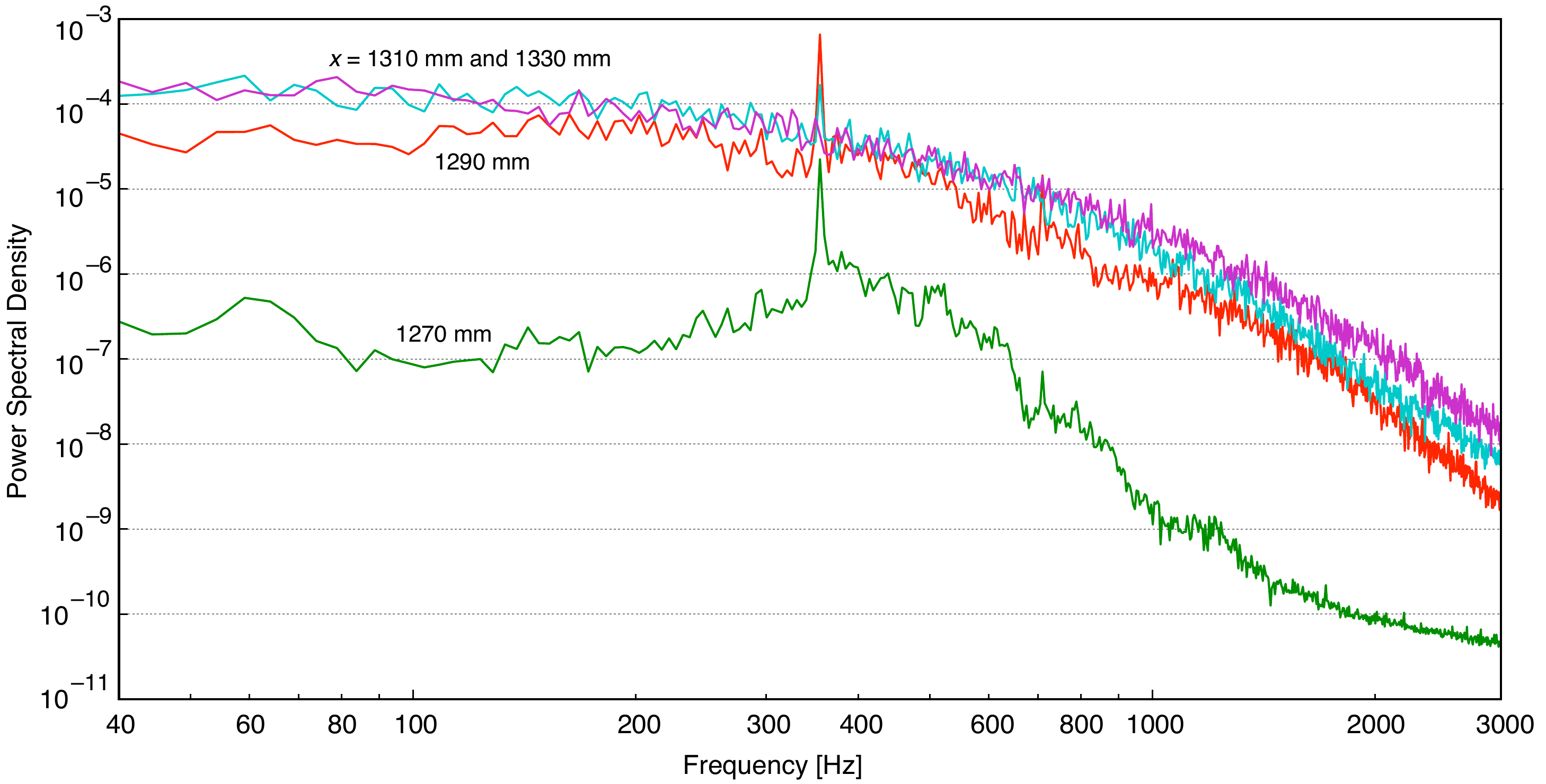}
    \caption{Velocity fluctuation power spectra in the initial low-speed streak for $\Rekk = 979$. Spectra are color-coded corresponding to the locations indicated in figure~\ref{fig:loc3e}.}
    \label{fig:Rek979PSDPSdudy1} 
\end{figure*}

\begin{figure}
    \centering \includegraphics[width=0.40\textwidth]{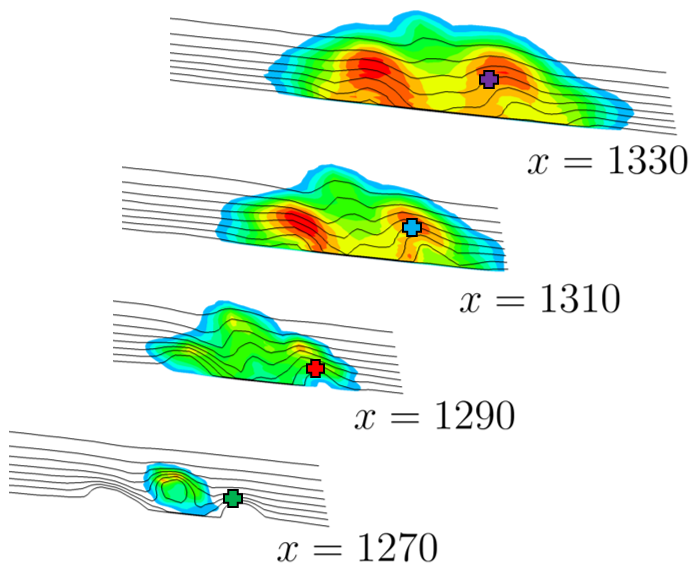}
    \caption{Locations of power spectra shown in figure~\ref{fig:Rek979PSDPSdudy1}.} 
    \label{fig:loc3e}
\end{figure}

Similar behavior is observed in the low-speed streaks that form at $\Rekk=750$ and~979. Data for $\Rekk=979$ is shown in figure~\ref{fig:Rek979PSDPSdudy1} for the locations indicated in
figure~\ref{fig:loc3e}. In this case, the unstable frequency band extends from around 250~to 700~Hz, $F=375$ to $900\times10^{-6}$. An important difference is the substantially increased fluctuation power at frequencies lower than the unstable band. This is thought to be a consequence of turbulence arising first on the $z=0$ centerline that has already occurred by $x=1270$~mm, the most upstream position pictured. Moving downstream, {transition} of the low-speed streak happens in the same manner as for $\Rekk=600$.

\begin{figure*}
    \centering \includegraphics[width=0.85\textwidth]{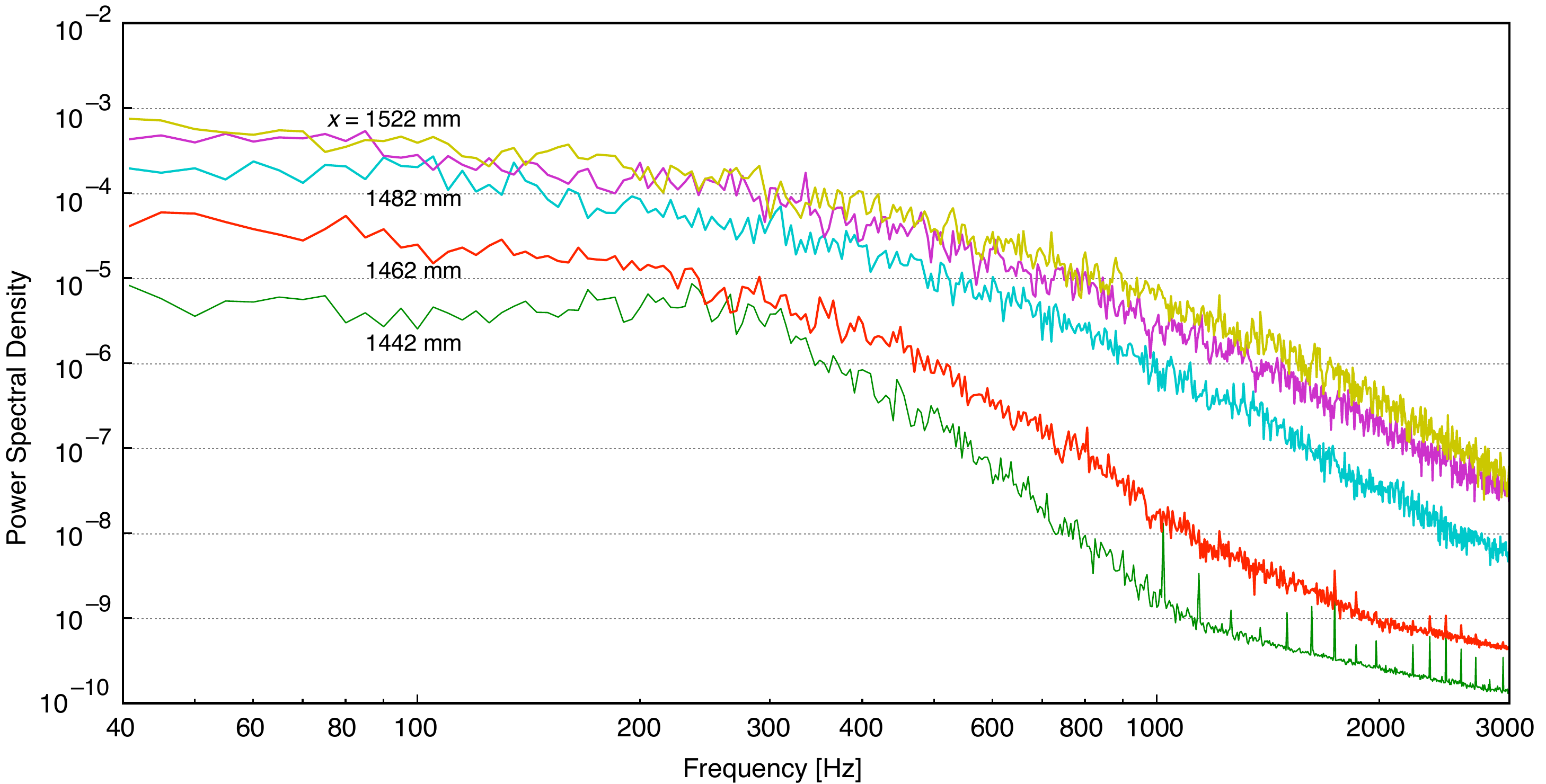}
    \caption{Velocity fluctuation power spectra in the second low-speed streak for $\Rekk = 600$. Spectra are color-coded corresponding to the locations indicated in figure~\ref{Rek600_S2_loc_edited}.}
    \label{Rek600PSD2ndStreak} 
\end{figure*}

\begin{figure}    
    \centering \includegraphics[width=0.45\textwidth]{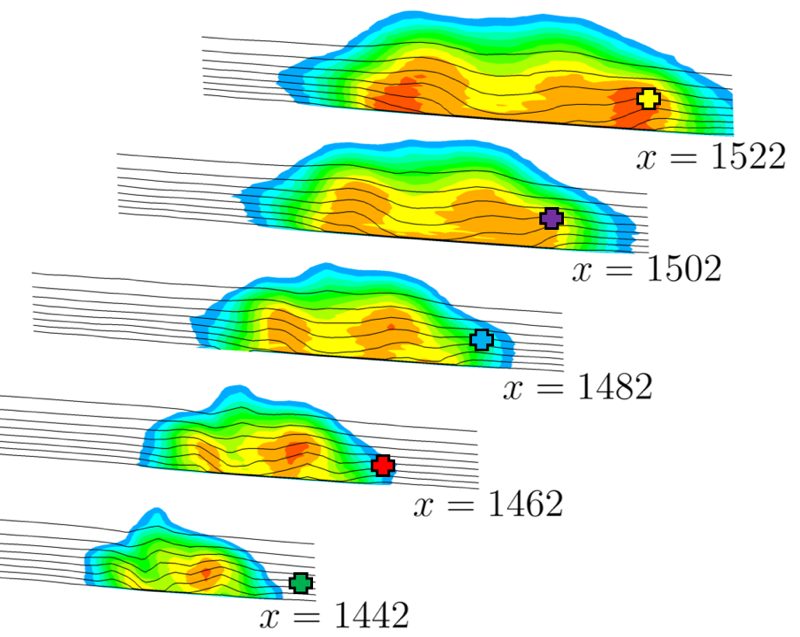}
    \caption{Locations of power spectra shown in figure~\ref{Rek600PSD2ndStreak}.} \label{Rek600_S2_loc_edited}
\end{figure}

Fluctuation spectra and corresponding measurement locations for the second $\Rekk=600$ low-speed streak are presented in figures~\ref{Rek600PSD2ndStreak} and~\ref{Rek600_S2_loc_edited}. In both cases, the high-intensity fluctuations of the turbulent wedge obscure any unstable frequency bands that may exist at those locations. Nevertheless, regions of distinctly stronger $u'$ fluctuations exist above the low-speed streaks as compared to other locations.

\begin{figure}
    \centering
    \includegraphics[width=0.5\textwidth]{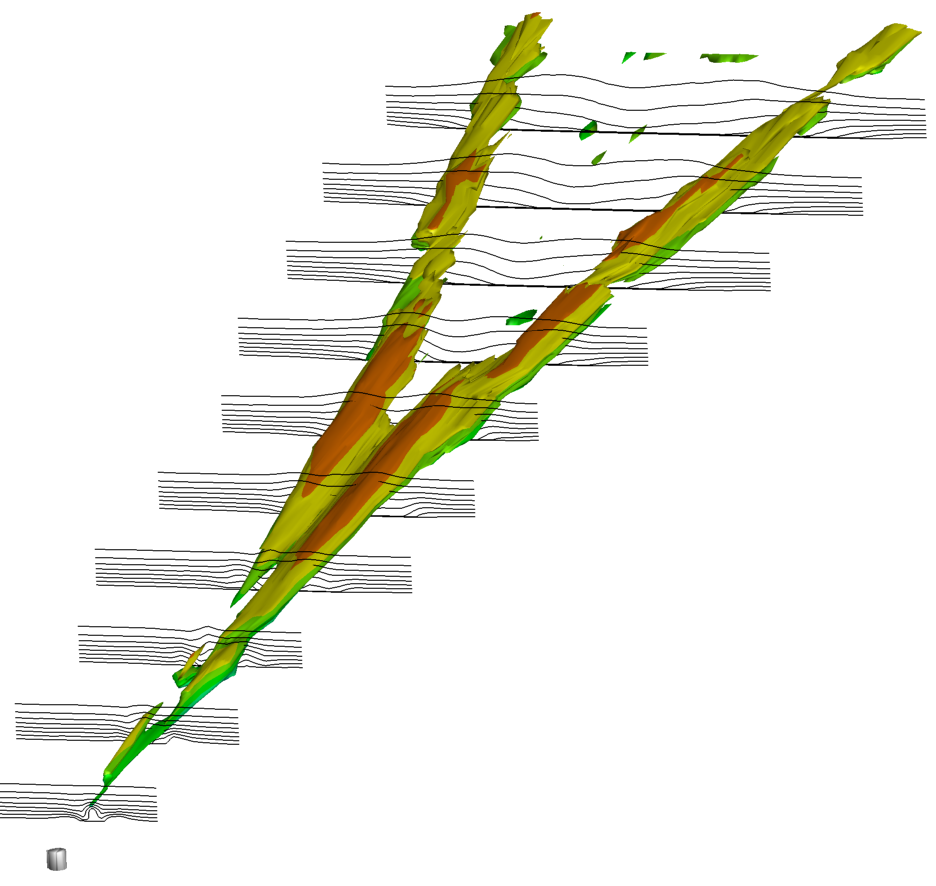}
    \caption{Iso-surfaces of $u'$ fluctuation power in the streak-instability bandpass (arbitrary level) for $\Rekk = 600$.} 
    \label{Rek600-Bandpass}
\end{figure}

Working from the assumption that regions of strongest $u'$ fluctuations may reveal the location of low-speed streaks, the fluctuation power spectrum data are integrated across the unstable frequency band identified in the initial low-speed streak. Iso-surfaces of the integrated power are plotted to show zones of highest-intensity fluctuations in the wedge. Fluctuation iso-surfaces of the $\Rekk=600$ data is presented in figure~\ref{Rek600-Bandpass}.

Figure~\ref{Rek600-Bandpass} recalls DNS results by Goldstein and co-workers \citep{Chu-AIAA-12,Goldstein-FTC-17}. Those simulations show that wedge spreading is driven by activity along wedge edges even when the core turbulence is artificially damped. The current observation of highest-intensity fluctuations in the streak-instability band is consistent with that finding.

\section{Conclusion}

These experiments seek to elucidate the characteristics and mechanisms of roughness-induced transition and turbulent wedge spreading.  It does so using a large number of wedge realizations generated by supercritical roughness elements with $\Rekk=600$, 750, and 979, and $Re_{x,k}$ values from $231\times10^{3}$ to $1.505\times 10^6$. The $Re_{x,k}$ range is much wider than earlier work and measurements are conducted over a substantially longer measurement domain.

Consistent with existing results, the naphthalene survey finds a mean spreading angle of $5.8^{\circ}$ for the turbulent core of the wedges. While there is significant data scatter, there is a sufficiently large sample size to suggest the wedge spreading angle increases weakly with $Re_{x,k}$, approximately $0.5^{\circ}$ per million Reynolds number across the range investigated.

When made nondimensional using $Re'$, the distance between the roughness and virtual wedge origin, \xo, does not exhibit a clear trend relative to $Re_{x,k}$ but does depend on \Rekk. The larger \Rekk\ values, 750 and 979, gave equivalent results, $\xo = 27\times10^3$, while the mean for $Re_{kk}=600$ is $74\times10^3$. The difference is explained by hotwire measurements that show initial breakdown to turbulence occurs high in the boundary layer just aft of the roughness element for the larger \Rekk\ configurations. At $\Rekk=600$, breakdown occurs first above the low-speed flanking streaks, further downstream. That is, even for supercritical \Rekk\ values, there are different types of transition and which type occurs affects the offset between the roughness element and the virtual wedge origin. These results reveal the underlying dynamics of the \Rekk\ versus $x$-based transition Reynolds number plots such as figure~10 by \citet{Klebanoff-JFM-92}.

Extensive hotwire measurements were conducted and found in good agreement with the naphthalene visualization results. The measurements show that, regardless of whether breakdown occurs first along the roughness centerline at high \Rekk\ values, velocity fluctuations eventually grow most strongly above low-speed streaks located at the flanks of the wedge. The growth and breakdown of these fluctuations drives lateral spreading of turbulence and produces high-speed streaks downstream and outboard of initiating low-speed streaks.  Subsequently, a new low-speed streak is observed outboard of the high-speed streak and continues the wedge-spreading process.

Velocity spectra show $u'$ fluctuations in a high-frequency band above the most-upstream low-speed streak. Spectra obtained on subsequent streaks are more broadband. It is not clear why the spectra are broadband in the more downstream low-speed streaks, even at relatively low fluctuation intensities. There are several possibilities.  (1)~The nearby turbulent flow in the inner part of the wedge affects the nearby hotwire measurements by their strong fluctuations.  (2)~These regions are truly turbulent, at least intermittently, with highest-intensity $u'$ fluctuations in the same upper shear layers as observed in the initial set of laminar low-speed streaks.  Or, (3)~wedge meandering is sufficiently strong that measurements above low-speed streaks are actually time averages of laminar, unstable and fully turbulent zones. However, preliminary evidence of meandering by \citet{Berger-PhD-20} suggests that meandering is most pronounced far downstream and is unlikely to completely obscure unstable bands in the second set of low-speed streaks.

Continuing to resolve the cycle of low- and high-speed streaks that drives turbulent wedge spreading remains an important problem in fundamental fluid mechanics. The laminar or turbulent character of low-speed flanking streaks is a key open question. The possibility of lateral meandering complicates progress on these issues. More sophisticated conditional averaging, perhaps using a reference sensor could help reveal modal behavior in order to successfully resolve these questions.

\bmhead{Acknowledgements}

The authors thank David Goldstein and Saikishan Suryanarayanan for fruitful discussions as well as Colton Finke for his assistance during the experiments. This work was funded by the United States Air Force Office of Scientific Research through grant FA9550-15-1-0345.

\section*{Declarations}

\textbf{Conflict of interest} The authors declare no competing interests.

\noindent\textbf{Ethical approval} Not applicable

\bibliography{sn-article}

\end{document}